# Effect of Input-output Randomness on Gameplay Satisfaction in Collectable Card Games

Yiwen Zhang[1], Diego Monteiro[*2†], Hai-Ning Liang[*2], Jieming Ma[2], Nilufar Baghaei[3]

Department of Computing
Xi'an Jiaotong-Liverpool University
Suzhou, China
[1]yiwenzhang1024@163.com,
[2]{ D.Monteiro, HaiNing.Liang, Jieming.Ma}@ xjtlu.edu.cn;
[*] Corresponding author

[†]DMT Lab
Birmingham City University
Birmingham, UK
Diego.VilelaMonteiro@bcu.ac.uk;

[3]School of Natural and
Computational Sciences
Massey University
N.Baghaei@massey.ac.nz

*Abstract*—Randomness is an important factor in games, so much so that some games rely almost purely on it for its outcomes and increase players' engagement with them. However, randomness can affect the game experience depending on when it occurs in a game, altering the chances of planning for a player. In this paper, we refer to it as "input-output randomness". Input-output randomness is a cornerstone of collectable card games like Hearthstone, in which cards are drawn randomly (input randomness) and have random effects when played (output randomness). While the topic might have been commonly discussed by game designers and be present in many games, few empirical studies have been performed to evaluate the effects of these different kinds of randomness on the players' satisfaction. This research investigates the effects of input-output randomness on collectable card games across four input-output randomness conditions. We have developed our own collectable card game and experimented with the different kinds of randomness with the game. Our results suggest that input randomness can significantly impact game satisfaction negatively. Overall, our results present helpful considerations on how and when to apply randomness in game design when aiming for players' satisfaction.

## I. Introduction

Randomness is an important factor in games; some games rely almost purely on it for its outcomes and to increase players' engagement level. In digital games, however, randomness can have other roles. It can enrich the gameplay experience, making it more diverse and unpredictable, an essential element of meaningful gameplay [1], [2]. Unpredicted information forces players to react and replan continuously. On the other hand, *when* unpredictability adaptation and replanning take place is a factor that can influence player enjoyment heavily. Due to its relevance and importance, when randomness happens according to its timing determines the name it receives.

*Input randomness* is the random information or element brought into the game before players make a decision [3] (see ). In contrast, *output randomness* is the random element brought into the game after players make a decision [3] (see ). The typical example of input randomness is when cards are drawn from a deck; the players are unaware of which cards they will have, but they can plan and decide what to do with the cards once they receive them. Output randomness can be exemplified by rolling the dice in *Risk*. The players choose how many troops to position and dice to roll, but they have no control over the outcome. In this paper, the term "input-output randomness" is the general notion of these kinds of randomness.

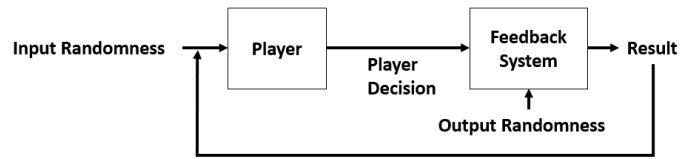

Fig. 1. An illustration of the continuous input-output randomness feedback loop in turn-based games.

While input-output randomness is a significant element in games, very few and limited studies have been done to understand it and explore its effect on gameplay and game satisfaction. The exploration of the effect of input-output randomness on game experience can provide a theoretical framework to allow deeper analysis and discussion

In this paper, we explore how input-output randomness affects several aspects of gameplay in a formal context. As cards are one of the most quintessential games of randomness, we developed our own collectable card game to investigate input-output randomness. We ran a controlled study and observed what aspects of the game experience were affected by the distinct kinds of randomness.

## II. Theoretical Foundation and Related Work

### A. Randomness

Video games and games of chance often have remarkably similar characteristics in that they both provide elements of randomness and intermittent rewards [4]. Randomness is not accepted equally. A study has shown that many people like to have the "illusion of control" [5]. This behaviour is apparent in studies involving gambling, where people feel they have better odds when using their own number instead of a machine-generated one [5].

Game designer Geoff Engelstein suggested that input randomness supports strategy, whereas output randomness undercuts players' strategic planning [6]. He argues that input randomness gives players the chance to assess the information and make plans, whereas output randomness may interrupt these plans and inject noise. Another game designer, Keith Burgun, agrees with this view and argues that "input randomness is definitely better than output randomness" [3]. For him, output randomness cuts off the correlation between game states, breaks up strategy or planning, makes no increase in the depth of a game, and obscures the game output (e.g., a player may have played perfectly but still lose the game) [7].

From these perspectives, output randomness takes a more significant role in players' level of anger and frustration than input randomness. However, Mark Brown notes that carefully tuned output randomness can improve a game, while poorly designed input randomness can damage the game [8]. Output randomness can be a tool to simulate mistakes and inaccuracies, forcing the player to consider risk management, which can work as future input randomness.

The above discussion and viewpoints on randomness in games, in our view, somewhat contrast with the important concept of the illusion of control and taking risks (i.e., gambling). In this exploratory analysis, we will investigate their differences.

*B. Collectable Card Games*

Collectable card games are games in which players design their own card decks by selecting cards from a pool of cards and then use this deck to play against an opponent who has a deck of their own [9]. Some notable examples of collectable card games are *Hearthstone: Heroes of Warcraft*[1] by Blizzard Entertainment and *Yu-Gi-Oh!*[2] by Konami. Collectable card games, by their nature, contain a high level of randomness (or at least the impression of randomness). At the beginning of each player's turn, the player draws a random hand, i.e., *input randomness*. After playing a card, it might have random effects, which is *output randomness*.

Fig. 1 (on page 1) presents how input-output randomness can affect gameplay in a collectable card game. Because collectable card games are sequential and turn-based, the *output* from one turn becomes the *input* to the next. In this way, we have a continuous turn-based randomness feedback loop. This repetitive pattern makes such games the perfect testing grounds for our two hypotheses because, since randomness is expected in these games, it can be manipulated and included with some degree of precision.

### III. EXPERIMENTAL DESIGN

We developed a card game and ran a user experiment with it. Fig. 2 shows a screenshot of the game, *Dream Cage*. The experiment had four Latin square counterbalanced conditions, and we used questionnaires to evaluate the game experience in each condition.

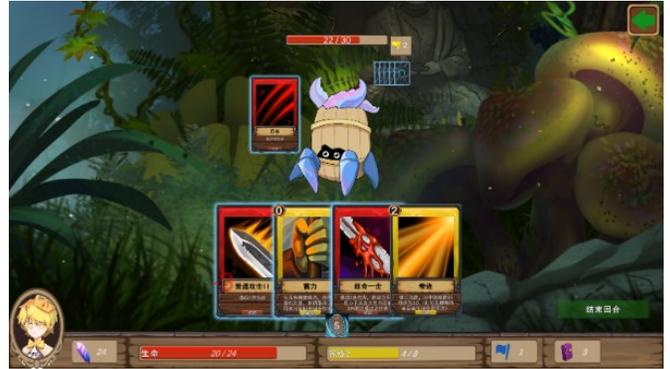

Fig. 2. Screenshots of *Dream Cage*, the card game that we developed and used to conduct our experiment. The picture is the battle screen, where players face opponents and must deplete the opponent's health bar by selecting which card(s) to play next. In the screenshot, the player lost four health points.

*A. Dream Cage*

*Dream Cage* was a typical collectable card game. That is, it followed the input-output format presented in Fig. 2. The game's goal was to defeat the opponents by selecting cards that would bring down the opponent's health points before depleting one's own health points. The health point information was displayed on a red health bar indicating how much damage the character took and was located on the lower middle half of the screen.

Before playing against an AI opponent, players were prompted to build their own deck from a pool of cards. The cards had attack descriptions of their expected behaviour written on them. To build the deck, players were given a budget in diamonds (purple rocks in the lower-left corner of Fig. 2). With this budget, players could choose from the available cards, and upon winning a battle, they would receive a prize in diamonds to reinforce their decks for the next battle. The game environment was written entirely in the local language (Chinese) to guarantee that players would not get confused because of linguistic barriers.

At the beginning of the battle match, players received five cards from the top of the deck and received an extra card in each subsequent turn. There was no limit to the number of cards players could hold during the game. This approach was used to avoid adding another mechanic noise that could influence the results. In the end, the player whose health bar reached zero first was considered the defeated one, and the game would stop.

Adversaries were presented as monsters and had their cards facing away from the player, which means the player was aware of how many cards the opponent had but not what the cards were (see Fig. 2, lower part). The opponent's card was only revealed upon being played. The player could simultaneously see only a few battle information of the adversary (i.e., health points and the number of cards).

The game was controlled using a standard mouse and clicking on the cards. Players could quit the game at any moment if they wished.

---

[1] https://playhearthstone.com/en-us.

[2] https://www.yugioh-card.com/en/.

*B. Experiment Conditions*

Following the input-output randomness framework presented earlier, our experiment had four conditions:

- *Input randomness* (*IR*). Players drew a certain number of cards, including "mystery" cards (i.e., random cards that the player had not previously added to their deck). The cards' effects were taken at face value without chance elements;

- *Output randomness* (*OR*). Players drew a certain number of cards from their deck without any *mystery cards* (i.e., all cards had been chosen by the player). Some cards produced unexpected results, and their effects could not be taken at face value;

- *Input and output randomness* (*IOR*). Players drew a certain number of cards, including *mystery cards* (similar to the IR condition). Like the OR condition, some cards produced unexpected results, and their effects could not be taken at face value; and

- *No input and output randomness* (*NR*). Players drew a certain number of cards from their deck without *mystery cards*. The cards' effects are taken at face value without chance elements.

The game and the four conditions were experimentally validated in a pilot study with a small set of 4 players. Each experimental condition lasts approximately 15 minutes. During the experiment, all conditions were counterbalanced to avoid carry-over learning effects.

The goal of the mystery card was to add a random element in the player's input beyond the one the player was expecting (i.e., a shuffled deck).

*C. Measurements*

We chose to use the GUESS [10] questionnaire to measure user experience. Following its guidance and our game genre, we selected *Usability*, *Play Engrossment*, *Enjoyment*, and *Creativity Freedom* as the primary subscales to form our questionnaire. The subscales were divided into items that, when averaged, present the score of their respective subscale. We calculated game satisfaction as the sum of the score of each subscale.

IV. EXPERIMENT

*A. Participants*

We recruited 18 college-age participants (6 females; 12 males) from a local university. They had an average age of 23.00 (s.d. = 1.71), ranging between 20 and 26. All volunteers had normal or corrected-to-normal vision, and none of them declared any health issues, physical or otherwise, that could have had an impact on the experiment.

*B. Procedure*

Due to Covid-19, the experiment was conducted online, following the guidelines and ethics procedures for running experiments. First, we explained the experiment's purpose to all participants and required them to download the game and read the experiment instructions. We also clarified any points about which the participants were not clear.

Each participant was allocated to one of the playing orders of the game modes in the experiment (balanced by a Latin square design). The trials were performed during four consecutive days, with participants playing one mode each day. After each interaction with the game, participants were requested to fill the GUESS questionnaire. The experiment for each day took approximately 20 minutes (15 minutes game + 5 minutes questionnaire) to finish. Fig. 4 shows the overall procedure of the experiment. On the last day, participants were allowed to freely express their opinions about the experiment and each mode.

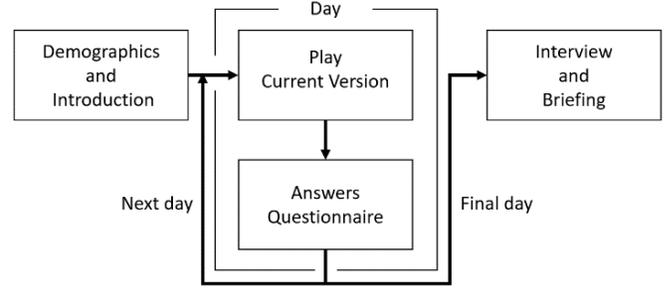

Fig. 3. Diagram of procedure of the experiment that took place over four consecutive days.

V. RESULTS

All subscales were submitted to a two-way repeated-measures analysis of variance (RM-ANOVA). The input-output randomness conditions were used to divide the data.

*A. Satisfaction and Subscales*

Satisfaction differed between the conditions with input randomness conditions (IR and IOR) and the ones without input randomness (NR and OR) $F (1, 15) = 6.275$, $p = 0.024$, $\eta_p^2 = 0.295$. Pairwise comparisons revealed that the satisfaction in the without input randomness (NR and OR) conditions (M = 22.991, 95% CI = [0.172, 2.137], p = 0.024) was larger than in the other conditions (IR and IOR) condition (M = 21.836, 95% CI = [20.844, 22.828]).

The analysis showed no significant effect of output randomness (OR and IOR) on game satisfaction when compared to the conditions without output randomness (NR and IR), $F (1, 15) = 0.33$, $p = 0.859$. The interaction of input randomness and output randomness (IOR) also did not produce a significant effect against (NR), $F (1, 15) = 1.112$, $p = 0.308$. The mean satisfaction did not differ between the conditions with output randomness (OR and IOR) (M = 22.459, CI = [21.103, 23.814]) and the ones without output randomness (NR and IR) (M = 22.368, CI = [21.489, 23.248]).

When explored further, IR had the lowest average ratings (M = 21.598, CI = [20.102, 23.094]) on satisfaction, which was statistically significant when compared to both NR and OR (both p < 0.05), but not statistically significant when compared to IOR (p > 0.05). The subscales did present significant differences between versions (p > 0.05).

*B. Subscales and Interview*

While many participants often abstained from making comments, some commented that they thought OR and NR conditions were more satisfying. They mentioned it because, in either condition, the game "*gives more place to strategic planning*", allowing them to make better action plans. One participant, however, pointed that these two conditions, OR and NR, made the game "*too easy*" because he could "*make a perfect plan*" to beat the AI opponent.

## VI. DISCUSSION

The results revealed that input randomness had a significant effect on game satisfaction in our collectable card game. The satisfaction in all conditions without input randomness was greater than with IR. However, it was not present in the IOR condition, suggesting that adding more than one kind of randomness into the system might produce noise and reduce the feeling of dissatisfaction from the positive aspect of input randomness. It is possible that the players had difficulty separating both kinds of randomness and conflated them, associating them with a game of pure chance. These results are in line with the findings from Goodman and Irwin [5], who found that the illusion of control is essential for enjoyment and meaningful gameplay. It is likely that randomness on its own was not pleasurable but not having "control over" the degree of randomness was a nuisance.

Even though the results are not in direct agreement with the opinions expressed by Brown [8], they can be reconciled with his discourse. The main point of his explanation of the importance of randomness and why one can be better accepted is controlling information and balancing players' agency and surprise. In our results, this was framed by the players who commented on the difficulty and their ability to plan. Furthermore, the participants' comments support this observation in that the ability to plan (or having "control" over game elements) made the participants more at ease.

Our study, however, did not find evidence to support that output randomness or the interaction of input and output randomness affect satisfaction significantly. Also, the effect of input-output randomness on the subfactors of Playability, Play Engrossment, Enjoyment, Creativity Freedom was not statistically significant.

*A. Limitations and Future Work*

We observed that although the statistical analysis did not show a significant effect of input randomness on subfactors ($p > 0.05$), which may indicate that our study was underpowered to explore these effects, they serve as a guide for further studies.

Moreover, in the experiment, we had four extreme input-output randomness conditions: input randomness, output randomness, both input and output randomness, and no randomness. We did not evaluate intermediary conditions, which could be the focus of future research.

Further investigations could address what aspects from input randomness were responsible for such results. Also, they could investigate other human factors that might have influenced the results because different personalities and gaming experiences might have contributed to our findings.

We evaluated a single kind of game (i.e., collectable card games) and, as such, it is unclear whether the results are translatable to other types of games. However, as cards and dice games are the quintessential examples of random games, it could likely be that the results could be transferable to other types of games, but further research is required to confirm this.

Finally, it will be interesting to explore the effects of input and output randomness with games that deal with more serious issues like mental health [11], where the primary goal is not just entertainment and whose players can be more sensitive to unpredictability and randomness.

## VII. CONCLUSION

In this research, we explored input-output randomness, a common and essential game mechanic used in a variety of games, including collectable card games. To our knowledge, the impact of input-output randomness on game satisfaction had not been studied in a formal context. We explored its effect on game satisfaction in a collectable card game based on four combinations of input-output randomness in a game we developed. Our experimental results revealed that input randomness significantly impacted game satisfaction in collectable card games. The game with input randomness was the least liked version in our experiment. We did not observe any statistically significant effect of output randomness and the interaction of input and output randomness on satisfaction levels. This research shines some light and opens further a discussion about the importance of randomness in games and how to best present it to players. Furthermore, it can be applied for the development of card games and as an alert on giving the players choice and removing it unexpectedly.

ACKNOWLEDGEMENT

The authors would like to thank the participants for their time and the reviewers for their reviews. The work is supported in part by Xi'an Jiaotong-Liverpool University (XJTLU) Key Special Fund (KSF-A-03; KSF-P-02), and XJTLU Research Development Fund.